\documentclass[]{piparticle-final}
\usepackage{graphicx}
\usepackage{amsmath}

\begin{document}

\volume{1}               % To be inserted by Editor
\articlenumber{010007}   % To be inserted by Editor
\journalyear{2009}       % To be inserted by Editor
\editor{A. C. Mart\'{\i}}   % To be inserted by Editor
%\reviewers{Reviewer's name}  % To be inserted by Editor
\received{14 October 2009}     % To be inserted by Editor
\accepted{28 December 2009}   % To be inserted by Editor
\runningauthor{A. D. Mariotti \itshape{et al.}}  % To be inserted by Editor
\doi{010007}         % To be inserted by Editor

\title{Parametric study of the interface behavior between two immiscible liquids flowing through a porous medium}

% Institution references with \cite are inserted after \maketitle in theaffiliation enviroment
\author{Alejandro David Mariotti,\cite{inst1}\thanks{E-mail: mariotti.david@gmail.com}  
        \hspace{1ex} Elena Brandaleze,\cite{inst2}\thanks{E-mail: ebrandaleze@frsn.utn.edu.ar}
        \hspace{1ex}Gustavo C. Buscaglia\cite{inst3}\thanks{E-mail: gustavo.buscaglia@icmc.usp.br}}

\pipabstract{
When two immiscible liquids that coexist inside a porous medium are drained through an opening, a complex flow takes place in which the interface between the liquids moves, tilts and bends. The interface profiles depend on the physical properties of the liquids and on the velocity at which they are extracted. If the drainage flow rate, the liquids volume fraction in the drainage flow and the physical properties of the liquids are known, the interface angle in the immediate vicinity of the outlet ($\theta$) can be determined. In this work, we define four nondimensional parameters that rule the fluid dynamical problem and, by means of a numerical parametric analysis, an equation to predict $\theta$ is developed. The equation is verified through several numerical assessments in which the parameters are modified simultaneously and arbitrarily. In addition, the qualitative influence of each nondimensional parameter on the interface shape is reported.
}

\maketitle

\blfootnote{
\begin{theaffiliation}{99}
   \institution{inst1} Instituto Balseiro, 8400 San Carlos de Bariloche, Argentina.
   \institution{inst2} Departamento de Metalurgia, Universidad Tecnol\'ogica Nacional Facultad Regional San Nicol\'as, 2900 San Nicol\'as, Argentina.
   \institution{inst3} Instituto de Ci\^encias Matem\'aticas e de Computa\c{c}\~ao, Universidade
de S\~{a}o Paulo, 13560-970 S\~{a}o Carlos, Brasil.
\end{theaffiliation}
}

\section{Introduction}

The fluid dynamics of the flow of two immiscible
liquids through a porous medium plays a key role in
several engineering processes. Usually, though the
interest is focused on the extraction of one of the
liquids, the simultaneous extraction of both liquids is
necessary. This is the case of oil production and of
ironmaking. The water injection method used in oil
production consists of injecting water back into the
reservoir, usually to increase pressure and thereby
stimulate production. Normally, just a small
percentage of the oil in a reservoir can be extracted,
but water injection increases that percentage and
maintains the production rate of the reservoir over a
longer period of time. The water displaces the oil from
the reservoir and pushes it towards an oil production
well [1]. In the steel industry, this multiphase
phenomenon occurs inside the blast furnace hearth, in which the porous medium consists of coke particles.
The slag and pig iron are stratified in the hearth and,
periodically, they are drained through a lateral orifice.
The understanding of this flow is crucial for the proper
design and management of the blast furnace hearth [2].
In both examples above, when the liquids are drained, a
complex flow takes place in which the interface
between the liquids moves, tilts and bends.

Numerical simulation of multiphase flows in porous
media is focused mainly in upscaling methods,
aimed at solving for large scale features of interest in
such a way as to model the effect of the small scale
features [3--5]. Other authors [6--8] use the
numerical methods to model the complex multiphase
flow that takes place at the pore scale.

In this work, we numerically study the macroscopic
behavior of the interface between two immiscible
liquids flowing through a porous medium when they
are drained through an opening. The effect of 
gravity on this phenomenon is considered. We define
four nondimensional parameters that rule the fluid dynamical
problem and, by means of a numerical
parametric analysis, an equation to predict the
interface tilt in the vicinity of the orifice ($\theta$) is
developed. The equation is verified through several
numerical cases where the parameters are varied
simultaneously and arbitrarily. In addition, the
qualitative influence of each non-dimensional
parameter on the interface shape is reported.

\section{Parametric Study}

The numerical studies in this work were carried out by
means of the program
FLUENT 6.3.26. Different models to simulate the
two-dimensional parametric study were used. The
volume of fluid (VOF) method was chosen to treat the
interface problem [9]. The drag force in the porous
medium was modeled by means of the source term
suggested by Forchheimer [10]. The source
term for the $i^{th}$ momentum equation is:

\begin{align}
S_i = -\left( \frac{\mu}{\alpha}V_i+\frac{1}{2}\rho C|\overrightarrow{V}|V_i\right). 
\end{align}

For the constants $\alpha$ and $C$ in Eq. (1), we use the
values proposed by Ergun [10]:

\begin{align}
\alpha = \frac{\varepsilon^3 d^2}{150(1-\varepsilon)^2}, 
\end{align}

\begin{align}
C = 1.75 \frac{(1-\varepsilon)}{\varepsilon^3 d}, 
\end{align}
where $\varepsilon$ is the porosity, $d$ is the particle equivalent diameter,
$\rho$ is the density, $V$ is the velocity and $\mu$ is the dynamic molecular viscosity.

Considering that the subscript 1 and 2 represent the
fluid 1 and the fluid 2 respectively, three nondimensional
parameters were considered in the
parametric study: viscosity ratio, $\mu_R=\mu_1/\mu_2$,
density ratio, $\rho_R=\rho_1/\rho_2$, and
nondimensional velocity, $V_R = V_0 \rho_2 L / \mu_2$; where
$V_0$ is the outlet velocity and $L$ a reference length.

\subsection{Domain description}

\begin{figure}%[th]
\begin{center}
\includegraphics[width=0.45\textwidth]{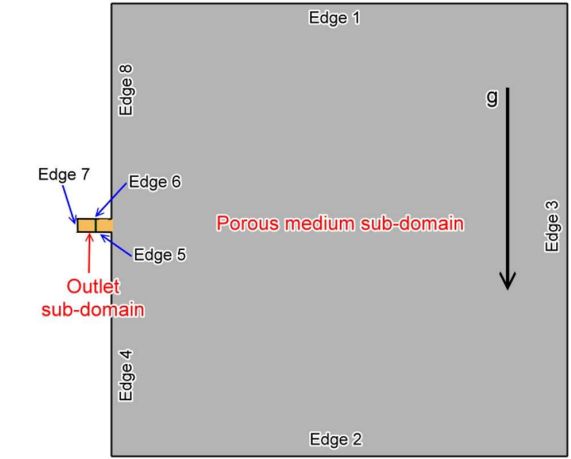}
\end{center}
\caption{Sketch of the numerical 2D domain.} \label{figure1}
\end{figure}

The numerical domain considered to carry out the
parametric study was a two-dimensional one
composed by the porous medium sub-domain and the
outlet sub-domain. The porous sub-domain is a
rectangle $10$m wide and $10$m tall. Inside of it, a rigid,
isotropic and homogeneous porous medium
was arranged. We use a porosity and particle diameter
of $0.32$ and $0.006$m, respectively.

For the outlet domain we use a rectangle $0.02 m$ wide
and with a height $L=0.01$m divided into two equal
parts and located at the center of one of the lateral
edges. The part located at the end of the outlet domain is
used to impose the outlet velocity. Figure 1 shows a
complete description of the domain.

A quadrilateral mesh with $2.2\times 10^4$ cells was used,
where the outlet sub-domain mesh consists of $200$
elements in all the cases studied.

As boundary conditions, on edge 1 we define a zero
gauge pressure condition normal to the boundary and
impose that only the fluid 1 can enter to the domain
through it. On edge 2 the boundary condition is the
same as on edge 1 but the fluid consider in this case is
fluid 2. On edge 7 we impose a zero gauge pressure
normal to the boundary but in this case the fluids can only leave
the domain. On the other edges (edges 4, 5, 6, and 8)
we impose a wall condition where the normal and
tangential velocity is zero except for edge 3, at which
the tangential velocity is free and the stress tangential
to the edge is zero.

\subsection{Interface evolution}

To illustrate how an interface reaches the stationary
position from an initially horizontal one, three sets of
curves were obtained.

\begin{figure}%[th]
\begin{center}
\includegraphics[width=0.45\textwidth]{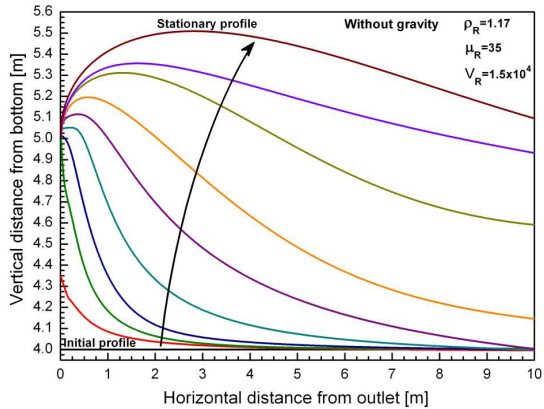}
\end{center}
\caption{Interface evolution when the initial position
is below the outlet level, without gravity.} \label{figure2}
\end{figure}

Figure 2 shows the interface evolution for the case
without the gravity effect and the interface initial
position is below the outlet level. The interface
modifies its tilt to reach the exit and it changes its
shape to reach the stationary profile.

Figures 3 and 4 show the interface evolution when the
gravity is present but the interface initial position is
below and above the outlet level respectively.

\begin{figure}%[th]
\begin{center}
\includegraphics[width=0.45\textwidth]{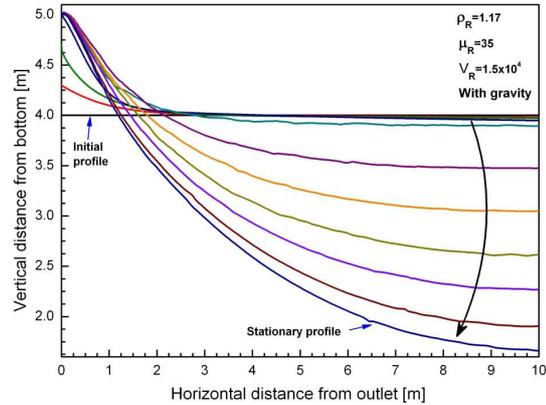}
\end{center}
\caption{Interface evolution when the initial position
is below the outlet, with gravity.} \label{figure3}
\end{figure}

\begin{figure}%[th]
\begin{center}
\includegraphics[width=0.45\textwidth]{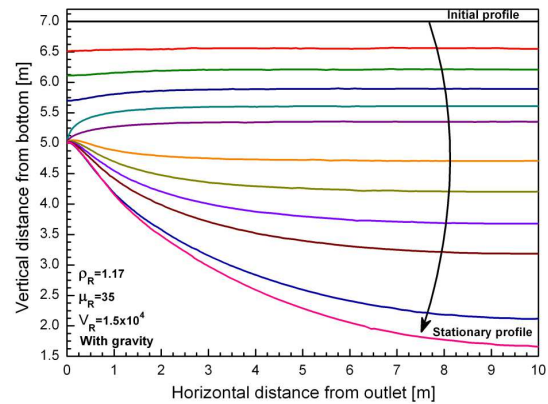}
\end{center}
\caption{Interface evolution when the initial position
is above the outlet, with gravity.} \label{figure4}
\end{figure}

\subsection{Viscosity effect}

One of the most important parameters to modify is the
dynamical viscosity of fluid 1. We maintain the
properties of the fluid 2 as the properties of water
(density $998$Kg/m$^3$, and dynamical viscosity $0.001$
Pa.s) and the density of fluid 1 as the density of the oil
($850$ Kg/m$^3$). The dynamical viscosity of fluid 1 was
varied from values smaller than those of fluid 2, to
values much greater.

\begin{figure}%[th]
\begin{center}
\includegraphics[width=0.45\textwidth]{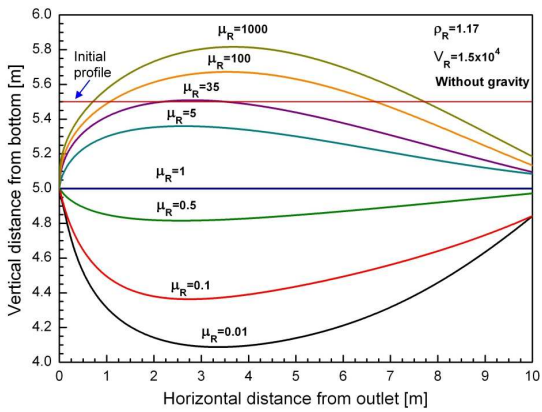}
\end{center}
\caption{Stationary interface profiles modifying the
fluid 1 viscosity without the gravity effect.} \label{figure5}
\end{figure}

Two sets of curves were obtained, one considering the
effect of gravity and the other without considering it.
Figure 5 shows the stationary interface profiles for the
different values of viscosity, without gravity. A value
of $V_R = 1.5\times 10^4$ and $\rho_R = 1.17$ were chosen. It is
possible to observe that, when fluid 1 has a viscosity
higher than that of fluid 2, the interface profile is
above the outlet and points downwards at the outlet. If
fluid 1 has a lower viscosity, the opposite happens.

\begin{figure}%[th]
\begin{center}
\includegraphics[width=0.45\textwidth]{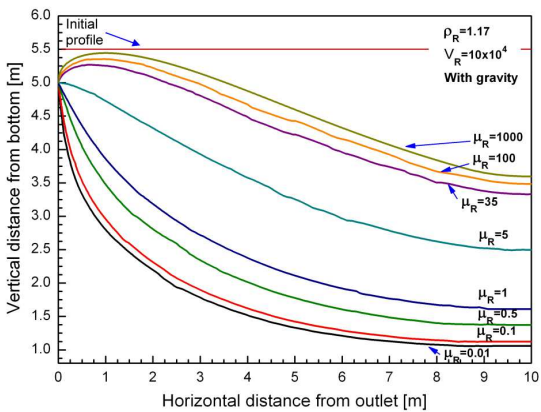}
\end{center}
\caption{Stationary interface profiles modifying the
fluid 1 viscosity with the gravity effect.} \label{figure6}
\end{figure}

When considering gravity, the value of $V_R$ was
changed to $1\times 10^5$ ($V_0 =10$m/s), since for smaller
values the interface may not reach the outlet (this is
later studied in Fig. 10). Figure 6 shows the curves
obtained for this situation, where the interface only lies
over the outlet level for the higher $\mu_R$ values.

\subsection{Outlet velocity effect}

$V_0$ is varied from a small value, similar to the porous
medium velocity ($V_0 = 0.2$m/s or $V_R = 2000$), to a
very large one ($V_0 = 50$m/s or $V_R = 5\times 105$).
Maintaining the properties of fluid 2 similar to those of
water, two sets of curves were obtained ($\mu_R>1$ and
$\mu_R<1$), shown in Figs. 7 and 8, respectively.

\begin{figure}%[th]
\begin{center}
\includegraphics[width=0.45\textwidth]{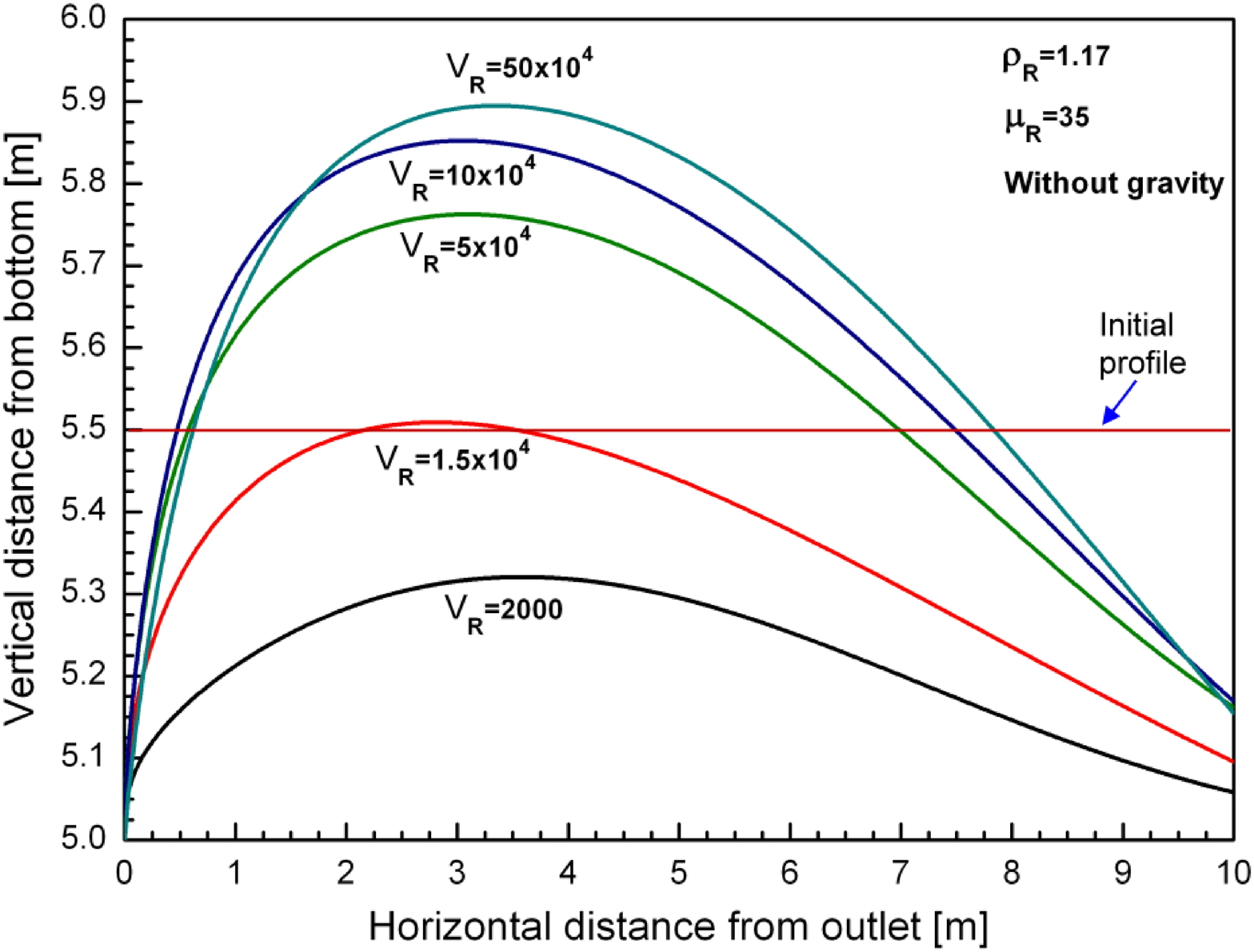}
\end{center}
\caption{Stationary interface profiles for several
values of $V_R$, without gravity, for $\mu_R>1$ ($\mu_R =35$).} \label{figure7}
\end{figure}

When the effect of gravity was considered, two
additional sets of curves (Figs. 9 and 10) were
obtained.

Figure 7 shows the effect of $V_R$ when the viscosity of
fluid 1 is greater than that of fluid 2, without gravity. It
is possible to see that, as $V_R$ increases, the interface tilt
at the outlet is maximal for $V_R=1\times 10^5$.

On the other hand, Fig. 8 shows the interface profiles
when the viscosity of fluid 1 is smaller than that of
fluid 2. We observe that as $V_R$ increases the interface
tends to the horizontal position.

\begin{figure}%[th]
\begin{center}
\includegraphics[width=0.45\textwidth]{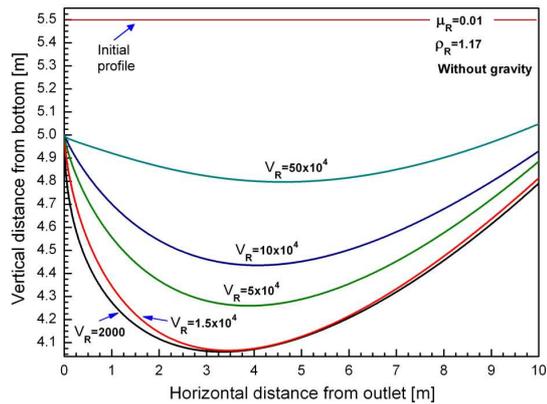}
\end{center}
\caption{Stationary interface profiles for several
values of $V_R$, without gravity, for $\mu_R<1$ ($\mu_R =0.01$).} \label{figure8}
\end{figure}

Figure 9 shows the different stationary interface
positions when the gravity effect is present for $\mu_R>1$.
The effect of gravity is quite significant, the interface
ascends but only for the highest value of $V_R$ it lies
above the outlet level.

\begin{figure}%[th]
\begin{center}
\includegraphics[width=0.45\textwidth]{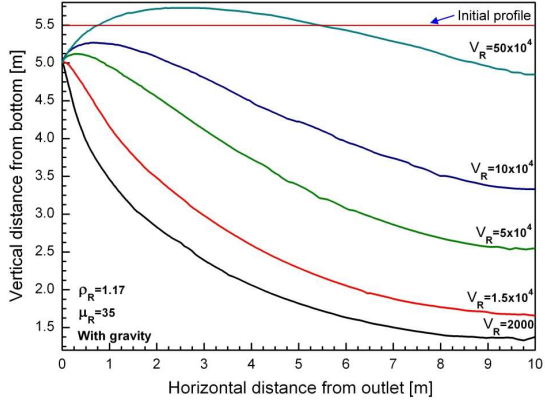}
\end{center}
\caption{Stationary interface profiles for several
values of $V_R$, with gravity, for $\mu_R>1$ ($\mu_R =35$).} \label{figure9}
\end{figure}

Figure 10 shows the curves when the viscosity of fluid
1 is lower than that of fluid 2 ($\mu_R<1$). The behavior is
different from that without gravity. In fact, there exists
a minimum outlet velocity below which the interface
does not reach the outlet.

\begin{figure}%[th]
\begin{center}
\includegraphics[width=0.45\textwidth]{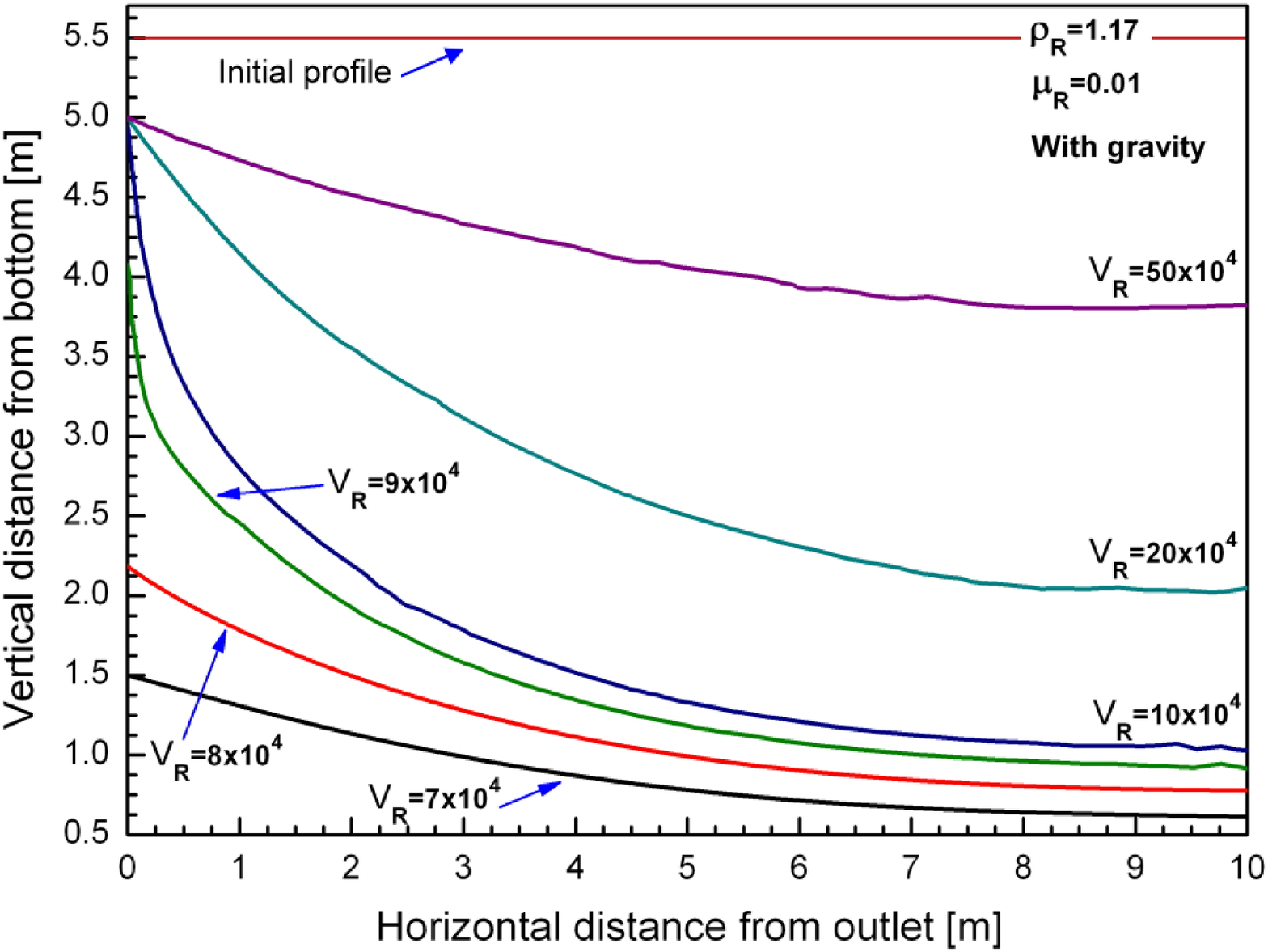}
\end{center}
\caption{Stationary interface profiles for several
values of $V_R$, with gravity, for $\mu_R<1$ ($\mu_R =0.01$).} \label{figure10}
\end{figure}

\subsection{Density effect}

The effect of the density ratio on the interface profile
was studied in the presence of gravity. Keeping fluid 2
with the properties of water and the viscosity of fluid 1
as $0.035$Pa.s ($\mu_R=35$), the density of fluid 1 was varied
from its original value to one three times smaller than
that of fluid 2. In Fig. 11 it is seen that as $\rho_R$
increases, the interface profile ascends significantly,
with a less significant change in the tilt angle at the
outlet.

\begin{figure}%[th]
\begin{center}
\includegraphics[width=0.45\textwidth]{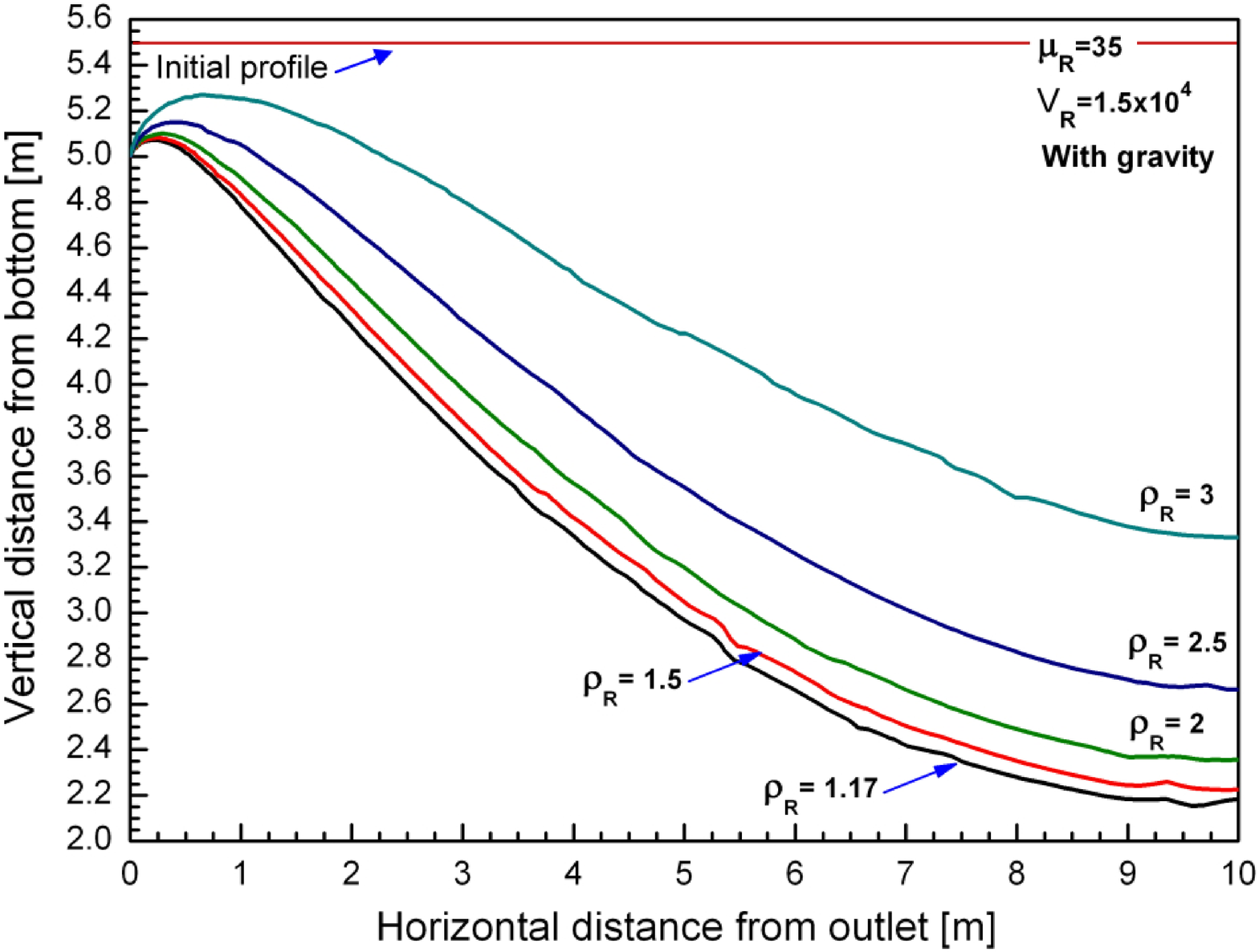}
\end{center}
\caption{Stationary interface profiles for several
values of the density of fluid 1, with $V_R=1.5\time 10^4$ and
$\mu_R=35$.} \label{figure11}
\end{figure}

\section{Generic expression}

From the study on the influence of each nondimensional
parameter on the interface behavior, an
equation that predicts the interface angle at the
immediate vicinity of the outlet ($\theta$) was crafted. For
practical reasons, the cases where gravity is present
were considered to develop the equation.

In Sect. 2.2, it is possible to see that when the nondimensional
parameters $\rho_R$, $\mu_R$ and $V_R$ are constant the
interface changes its shape until it reaches a stationary
profile.

For this reason, a fourth nondimensional parameter is
considered, the volume fraction of fluid 1 in the outlet
flow (VF).

A generic expression [(Eq. (4)], consisting of three
terms and containing 22 constants, was adjusted by trial
and error until satisfactory agreement with the
numerical results was found.

\begin{align}
\theta &= \alpha \mu_R^b V_R^c \rho_R^d \notag \\
&+ e \mu_R^f V_R^g \rho_R^h \exp(-i \mu_R^j V_R^k \rho_R^l VF^m) \notag \\
&+ n \mu_R^o V_R^p \rho_R^q \exp(-r \mu_R^s V_R^t \rho_R^u VF^v) 
\end{align}

Table 1 shows the values of the constants in the
generic expression.

\begin{table}
\begin{center}
\begin{tabular}{cc|cc|cc}
\hline\hline
a & -29.1 & i & $1.162\times 10^7$ & p & 0.1 \\
 b & -0.04 & j & -0.18 & q & 0.72 \\
 c & 0.1 & k & -1.64 & r & 2763.1 \\
 d & 0.45 & l & -1 & s & 0.4 \\
 e & 206 & m & 0.63 & t & -0.6 \\
 f & 0.035 & n & 12.74 & u & -1.82 \\
 g & -0.05 & o & 0.085 & v & 3.2 \\
 h & -1.26 & & & & \\
\hline\hline
\end{tabular} 
\end{center}
\caption{Constant values in the generic expression.}
\end{table}

\subsection{Equation verification}

To verify that the generic expression (4) predicts the
value of $\theta$ correctly when the parameters are arbitrarily
modified, 21 additional numerical cases were
simulated. These cases cover a wide range of physical
properties of the liquids and of the characteristics of
the porous medium (by means of the coefficients $1/\alpha$
and $C$).

The interface angle obtained from the generic
expression ($\theta_{GE}$) was compared with the interface
angle obtained from the simulations ($\theta_S$).
Table 2 shows the five porous medium types (PT) that
were chosen.

\begin{table}
\begin{scriptsize}
\begin{center}
\begin{tabular}{p{7pt}|ccccc}
\hline\hline\\
PT & $D$ & $\varepsilon$ & $1/\alpha$ & $C$ & Resistance \\
\hline A & 0.005 & 0.3 & $10.88\times 10^7$ & $1.81\times 10^4$ & Very high \\
 B & 0.006 & 0.32 & $5.88\times 10^7$ & $1.21\times 10^4$ & High \\
 C & 0.02 & 0.2 & $3\times 10^7$ & $1.75\times 10^4$ & Medium \\
 D & 0.02 & 0.25 & $1.35\times 10^7$ & 8400 & Low \\
 E & 0.05 & 0.17 & $8.41\times 10^6$ & $1.18\times 10^4$ & Very low \\
\hline\hline
\end{tabular} 
\end{center}
\end{scriptsize}
\caption{Porous medium types.}
\end{table}

Some cases (1, 2, 4, 5, 9, 11, 17-21) were chosen
based on the possible combinations of immiscible
liquids that can be manipulated in real situations. For the
remaining cases, the properties of the liquids were
fixed at arbitrary values (fictitious liquids) so that a
wide range of the nondimensional parameters was
covered. Table 3 shows the description of each
numerical case, while Table 4 shows the
corresponding nondimensional parameter values and
PT.

\begin{table}
\begin{scriptsize}
\begin{center}
\begin{tabular}{p{17pt}|p{21pt} p{21pt} cccc}
\hline\hline Case & \begin{tiny}Fluid 1\end{tiny} & \begin{tiny}Fluid 2\end{tiny} & $\mu_1$ & $\mu_2$ & $\rho_1$ & $\rho_2$ \\
\hline 1 & Heavy oil & Water solution & 0.4 & 0.005 & 850 & 998 \\
 2 & Light oil & Water emulsion & 0.012 & 0.06 & 850 & 998 \\
 3 & -- & -- & 0.08 & 0.008 & 4000 & 4680 \\
 4 & Kero-sene & Water & $24\times 10^{-4}$ & 0.001 & 780 & 998 \\
 5 & Ace-tone & Water & $3.3\times 10^{-4}$ & 0.001 & 791 & 998 \\
 6 & -- & -- & 0.003 & 0.01 & 400 & 720 \\
 7 & -- & -- & 0.2 & 0.01 & 1500 & 2700 \\
 8 & -- & -- & 0.3 & 0.004 & 3300 & 5940 \\
 9 & Light slag & Hot pig iron & 0.02 & 0.001 & 2800 & 7000 \\
 10 & -- & -- & 0.5 & 0.013 & 500 & 1250 \\
 11 & Medium & pig & 0.4 & 0.005 & 2800 & 7000 \\
         & slag & iron & & & & \\
 \begin{tiny}12-16\end{tiny} & -- & -- & 0.08 & 0.008 & 4000 & 4680 \\
 \begin{tiny}17-21\end{tiny} & Heavy slag & pig iron & 0.4 & 0.005 & 2800 & 7000 \\
\hline\hline 
\end{tabular} 
\end{center}
\end{scriptsize}
\caption{Cases description.}
\end{table}

\begin{table}
\begin{small}
\begin{center}
\begin{tabular}{c|ccccc}
\hline\hline Case & PT & $\rho_R$ & $\mu_R$ & $V_R$ & VF \\
\hline 1 & B & 1.17 & 80 & $5\times 10^4$ & 2.8 \\
 2 & B & 1.17 & 0.2 & $5\times 10^4$ & 87.2 \\
 3 & B & 1.17 & 10 & $10\times 10^4$ & 28.4 \\
 4 & B & 1.28 & 2.4 & $5\times 10^4$ & 96.1 \\
 5 & B & 1.26 & 0.33 & $1\times 10^5$ & 96.8 \\
 6 & B & 1.8 & 0.3 & $1\times 10^5$ & 93.2 \\
 7 & B & 1.8 & 20 & $1\times 10^5$ & 16.1 \\
 8 & B & 1.8 & 80 & $8\times 10^4$ & 18.4 \\
 9 & B & 2.5 & 20 & $1\times 10^5$ & 100.0 \\
 10 & B & 2.5 & 40 & $1\times 10^5$ & 5.5 \\
 11 & B & 2.5 & 80 & $5\times 10^4$ & 70.8 \\
 12 & A & 1.2 & 10.0 & $1\times 10^5$ & 23.1 \\
 13 & B & 1.2 & 10.0 & $1\times 10^5$ & 28.4 \\
 14 & C & 1.2 & 10.0 & $1\times 10^5$ & 41.5 \\
 15 & D & 1.2 & 10.0 & $1\times 10^5$ & 53.4 \\
 16 & E & 1.2 & 10.0 & $1\times 10^5$ & 63.7 \\
 17 & A & 2.5 & 80.0 & $5\times 10^4$ & 15.8 \\
 18 & B & 2.5 & 80.0 & $5\times 10^4$ & 24.3 \\
 19 & C & 2.5 & 80.0 & $5\times 10^4$ & 43.2 \\
 20 & D & 2.5 & 80.0 & $5\times 10^4$ & 70.8 \\
 21 & E & 2.5 & 80.0 & $5\times 10^4$ & 94.0 \\
\hline\hline 
\end{tabular} 
\end{center}
\end{small}
\caption{Parameter values and PT for all cases described in Table 3.}
\end{table}

We define an error ($e = 100|\Delta\theta|/180$) as the percentage of the
absolute value of the difference between the interface
angles ($\Delta\theta = \theta_S - \theta_{GE}$) divided by the interface angle range
($180^{\circ}$). Figure 12 shows the comparison between the
generic expression and the numerical cases. It is seen
that the generic expression (4) predicts the interface
angle, for the cases used in this study, with an error
smaller than 10\%.

\begin{figure}%[th]
\begin{center}
\includegraphics[width=0.45\textwidth]{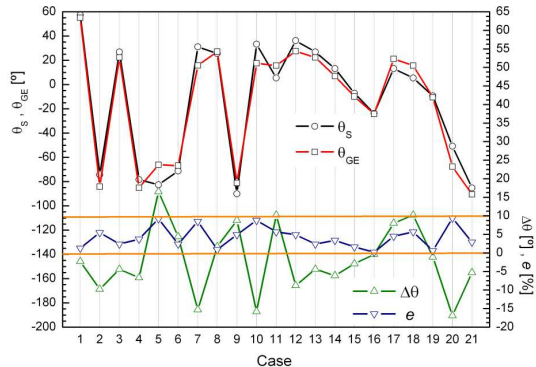}
\end{center}
\caption{Comparison between the predictions of the
generic expression and the numerical result for the 21
validation cases.} \label{figure12}
\end{figure}

\section{Conclusions}
A numerical study of the macroscopic interface
behavior between two immiscible liquids flowing
through a porous medium, when they are drained
through an opening, has been reported. Four nondimensional
parameters that rule the fluid-dynamical
problem were identified. Thereby, a numerical
parametric analysis was developed where the
qualitative observation of the resulting interface
profiles contributes to the understanding of the effect
of each parameter. In addition, a generic expression to
predict the interface angle in the immediate vicinity of
the outlet opening ($\theta$) was developed. To verify that
the generic equation predicts the value of $\theta$ correctly,
$21$ numerical cases with widely different parameters
were simulated. Considering that the cases encompass
a large class of liquids and porous media, the
prediction of $\theta$ within an error of 10\% is
considered satisfactory.

\begin{acknowledgements}
A. D. M. and E. B. are grateful for the support from
Metallurgical Department and DEYTEMA (UTNFSRN).
G. C. B. acknowledges partial financial support
from CNPq and FAPESP (Brazil).
\end{acknowledgements}

\end{document}